\documentstyle[floats,aps,epsf,prl]{revtex}

\begin{document}
\draft
\preprint{April 2, 1997}
\twocolumn[\hsize\textwidth\columnwidth\hsize\csname
@twocolumnfalse\endcsname
\title{A period-doubled structure for the $90^{\circ}$ partial
dislocation in silicon}
\author{J. Bennetto, R.W. Nunes, and David Vanderbilt}
\address{Department of Physics and Astronomy, Rutgers University,
Piscataway, New Jersey 08855-0849}

\date{April 2, 1997}
\maketitle

\begin{abstract}
We suggest that the commonly-accepted core structure of the 90$^\circ$
partial dislocation in Si may not be correct, and propose instead a
period-doubled structure.  We present LDA, tight-binding, and classical
Keating-model calculations, all of which indicate that the
period-doubled structure is lower in energy.  The new structure
displays a broken mirror symmetry in addition to the period doubling,
leading to a wide variety of possible soliton-like defects and kinks.
\end{abstract}
\vskip1pc]

\narrowtext

Dislocations in silicon and other semiconductors have been well
studied both theoretically and experimentally\cite{hl}.
They are well known to be responsible for plastic behavior,
and affect electronic properties as well.  The predominant
dislocations in silicon lie along the $\langle 110\rangle$
directions, within a $\{111\}$ slip plane, with Burgers vectors
at $0^{\circ}$ or $60^{\circ}$ to the propagation
direction.  These disassociate into partial dislocations separated
by a ribbon of stacking fault.  The 0$^\circ$ splits into two
$30^{\circ}$ partials, while the 60$^\circ$ splits into a
$30^{\circ}$ and a $90^{\circ}$ partial.

The core structure of the $90^{\circ}$ partial has received much
attention.  The unreconstructed core contains a zigzag chain of
three-fold coordinated atoms.  It has been proposed
\cite{hirsch79,jones79,markl79} that this dislocation core
reconstructs by breaking a $\{110\}$ mirror symmetry, as shown in
Figs.~1(a) and 2(a), in order to eliminate the dangling bonds.  Thus,
each under-coordinated atom forms a new bond with a partner on the other
side of the zigzag chain, and the defect core becomes fully saturated.
Several workers have shown theoretically \cite
{markl83,chel84,lodge89,bigger92,markl92,jones93,markl94,nunes96}
that this reconstruction lowers the energy by approximately 0.7
eV per unit cell, or 0.18 eV/\AA, with respect to the symmetric case.
This might be expected, as it restores the four-fold coordination of all
the atoms, albeit at the cost of some local bond strain.  Moreover, EPR
measurements find a low density of dangling bonds, supporting full
reconstruction\cite{hirsch85}. Thus, a consensus seems to have emerged
that this reconstruction represents the physically correct core
structure, and a large volume of work has come to rely on this
assumption\cite{jones80,heggie83,heggie93,hansen95}.

In this Letter, we propose a new structure for the core of the
90$^\circ$ partial dislocation in Si.  Our proposed structure
retains the four-fold coordination of every atom in the core,
but introduces a doubling of the periodicity along the dislocation
direction.  The new structure is found to be lower in energy
than the previously assumed reconstruction, regardless of whether
the comparison is based on empirical interatomic potential,
total-energy tight-binding, or first-principles density-functional
calculations.  Thus, it appears likely that all previous work on
the 90$^\circ$ partial has assumed an incorrect core structure, and
that the interpretation of experimental studies on this dislocation
system should be reexamined in light of the new structural model.

Our proposed, period-doubled structure is shown in Figs.~1(b) and 2(b).
We shall refer to it as the double-period (DP) structure, in
contrast to the single-period (SP) structure of Figs.~1(a) and 2(a).
The DP structure can be derived from the SP one by
inserting alternating kinks at every lattice site along the core.  This
shifts the center of the dislocation core by one-half lattice spacing
along the slip plane, so that the center of the DP core is located
halfway between neighboring possible positions of the SP core (Fig.~1).
Like the
SP structure, the DP one is built entirely out of 5-, 6-, and 7-fold
rings.  It also retains the symmetry breaking of the SP structure,
violating mirror symmetry across the (110) plane.  Thus, the DP core
has four equivalent ground states, related to each other by (110)
mirrors and by single-cell translations.  This makes for an especially
rich spectrum of solitonic defects and kinks, as we shall see below.

\begin{figure}
\epsfxsize=3.5 truein
\epsfbox{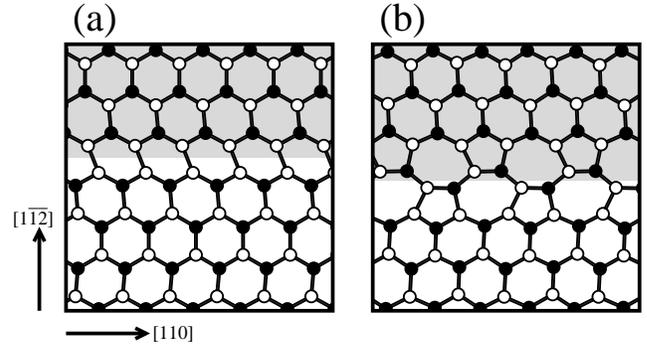}
\caption{(a) SP structure of the 90$^\circ$ partial, viewed from above
the $(1\bar11)$ slip plane.  Shaded region indicates stacking fault.
Black (white) atoms lie below (above) the slip plane. (b) Same
view of the DP structure.}
\end{figure}

We apply three different approaches to calculate the relative energies
of the SP and DP core structures.  First, we use the Keating
model\cite{keating}, a classical interatomic potential model containing
nearest-neighbor bond stretching and bending force constants.  Since
both core structures contain only four-fold Si atoms, the Keating
energies might be expected to give a reasonable first approximation.
Second, we use a total-energy tight-binding (TETB) approach, in which
the electrons are treated quantum-mechanically but in an empirical
framework.  This approach was implemented using the linear-scaling
density-matrix method of Li {\it et al.}\cite{lnv}, with a real-space
density-matrix cutoff of 7.33 \AA, and the electron chemical potential
in the middle of the band gap.  We used the tight-binding
parameterization of Kwon {\it et al.}\cite{kwon}.  Other details are as
in Ref.~\onlinecite{nunes96}.  Third, on system sizes up to about 200
atoms, we carried out {\it ab-initio} calculations within the
local-density approximation (LDA) to density-functional theory.  A
plane-wave pseudopotential approach was employed, using a
Kleinman-Bylander pseudopotential with $s$-nonlocality only
\cite{perez}, and a plane-wave cutoff of 7 Ry.  In all three cases,
forces were relaxed to better than $5\times 10^{-3}$eV/\AA\ per atom,
with an average force of less than $5\times 10^{-5}$eV/\AA.

\begin{figure}
\epsfxsize=2.0 truein
\centerline{\epsfbox{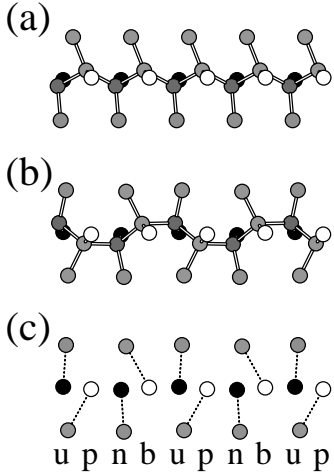}}
\caption{(a) SP structure (same view as in Fig.~1) but showing only
``core'' atoms and their neighbors.  Darker atoms are farther away.
(b) Same view of the DP structure.
(c) Schematic representation of (b), in which core atoms have
been removed, and second-neighbor connections between remaining
atoms are shown.  Corresponding symbolic notation is indicated
(see text).}
\end{figure}

For calculations on the SP structure, we have constructed supercells
containing 96 atoms (``smaller'' cell) or 288 atoms (``larger'' cell);
for the DP structures, these are doubled to 192 and 576 atoms,
respectively.  In terms of underlying lattice vectors
${\bf a}={a\over 2}[1\bar1\bar2]$,
${\bf b}={a\over 2}[110]$,
${\bf c}=a[1\bar11]$
representing a 12-atom orthorhombic cell, the 96-atom supercell is
defined as
${\bf a}'=4{\bf a}$,
${\bf b}'={\bf b}$,
${\bf c}'=2{\bf c}+2{\bf a}+{1\over6}{\bf a}$.
The ${\bf c}'$ vector is chosen to situate the dislocations in a
quadrupole lattice to avoid the spurious shear strains in the minimal
dipole cell\cite{bigger92}. (The extra term ${1\over6}{\bf a}$ in $\bf
c'$ relieves the strain introduced by the ribbon of stacking fault.)
The 192-atom cell has ${\bf b'}=2{\bf b}$, but is otherwise identical to
the 96-atom one.  The LDA calculations on these supercells were
performed with the two k-points (0,1/8,0), (0,3/8,0) for the DP
structure, and the corresponding 4-point set for the SP structure.
Empirical and tight-binding calculations were also carried out for
enlarged supercells of 288 (SP) or 576 (DP) atoms having lattice vectors
${\bf a'}=8{\bf a}$, ${\bf b'}={\bf b}$ or $2\bf b$, and ${\bf c}'=3{\bf
c}+4{\bf a}+{1\over6}{\bf a}$.

Table I shows the results of our total-energy calculations on these
cells.  For the case of the SP structure, we find that the total
energy of the supercell differs noticeably depending on whether
the direction of the mirror symmetry-breaking is the same, or
opposite, for the two dislocations in the supercell.  In the Table,
$\overline{E}_{\rm SP}$ refers to the average of these two
energies, while $\Delta E_{\rm SP}$ refers to the difference.  The
corresponding energy splitting is not significant in the DP case.
We expect $\overline{E}_{\rm SP} - E_{\rm DP}$ to be a reasonable
estimate of the relative energy of SP and DP dislocations in the
limit of large supercell size.  Note that in all cases $E_{\rm DP}$
is energetically favored not only over $\overline{E}_{\rm SP}$, but
also over the preferred of the two SP configurations.
In view of the Keating result, it appears likely that the DP
structure is preferred because it is able to reduce the local
bond strains near the core.  Probably this is associated with
the fact that the DP structure breaks the (110) mirror symmetry
more gently than does the SP one.

\begin{table}
\caption{Calculated energy differences between core reconstructions of
the 90$^\circ$ partial dislocation, in meV/\AA.  Cell size refers to the
double-period cell.  $E_{\rm DP}$ is the energy of the double-period
reconstruction.  For the single-period case, $\overline{E}_{\rm SP}$ and
$\Delta E_{\rm SP}$ are respectively the average and difference of the
energies for the two different relative arrangements of mirror
symmetry-breaking.}
\begin{tabular}{lcccc}
 &\multicolumn{2}{c}{192-atom supercell}
 &\multicolumn{2}{c}{588-atom supercell}\\
 &$E_{\rm DP} - \overline{E}_{\rm SP}$ &$\Delta E_{\rm SP}$
 &$E_{\rm DP} - \overline{E}_{\rm SP}$ &$\Delta E_{\rm SP}$\\
\hline
Keating		&-22	&19	&-10	& 4	\\
TETB            &-76	&39	&-55	& 8	\\
LDA		&-79	&47	\\
\end{tabular}
\end{table}

Clearly, our results suggest that the DP structure ought to be the
physically realized core structure for the 90$^\circ$ partial
dislocation in Si.  In view of the extensive experimental work
on this system, it seems surprising that such a possibility
should have been overlooked.  However, the two structures do
have much in common.  Both the SP and DP structures are fully
reconstructed, and thus neither gives rise to deep-gap states
that would be expected to show an ESR signal.  Both are
constructed entirely of 5-, 6-, and 7-fold rings,
and the maximally strained bonds show comparable distortions in
the two cases.  Thus, there does not appear to be any obvious
signature in electrical or optical properties that would
distinguish the DP from the SP structure.  Regarding imaging,
remarkable progress has been made with transmission electron
microscopy (TEM), to the point where individual kinks in the
30$^\circ$ and 90$^\circ$ partials can be resolved\cite{kolar96}.
While the proposed period doubling is not evident in the
90$^\circ$ core in these images, neither is it visible in the
core of the 30$^\circ$ partial, for which a DP structure is well
accepted.  Nor does it appear possible to locate the position
of the 90$^\circ$ core to a resolution of better than half a
lattice spacing, which also might distinguish between the SP
and DP structures.  Thus, it appears that the resolution of TEM
is still not adequate to settle this issue.  Previous calculations
of the activation energies for kink formation and migration in
the SP structure\cite{nunes96} were found to be in reasonable
($\sim$20\%) agreement with experiment, but this agreement
may have been fortuitous.

Thus, to our knowledge, present experiments neither rule out
nor support our identification of the DP structure as the correct
ground-state structure for the 90$^\circ$ partial.  It is to
be hoped that the present results will stimulate further
experimental investigations of this issue.  For example,
perhaps some kind of imaging electron diffraction technique
might be capable of observing the proposed period doubling
in the dislocation core.

\begin{figure}
\epsfxsize=3.5 truein
\epsfbox{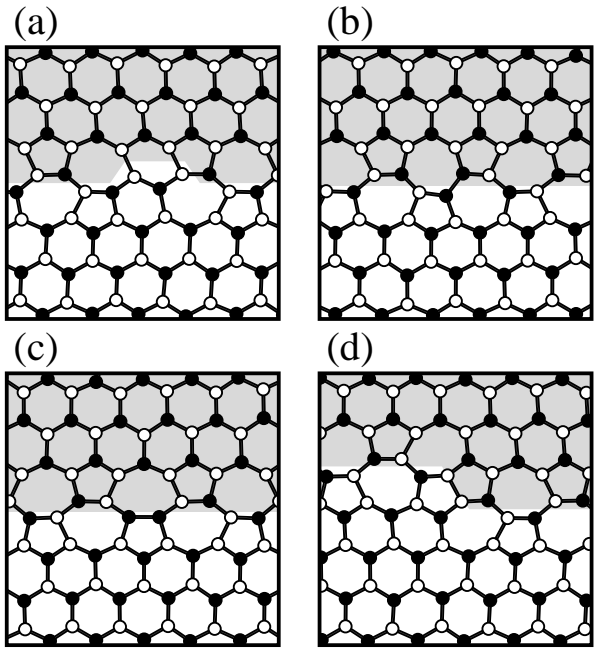}
\caption{Examples of several types of core defects in the DP
structure.  Viewpoint is the same as for Fig.~1.
(a) Phase-switching defect (PSD).
(b)-(c) Direction-switching defects (DSD). (d) Kink.}
\end{figure}

In the remainder of this Letter, we discuss the structural defects
that can occur for the DP core structure, including solitonic
and kink structures.  We first introduce a shorthand notation
for describing the possible core structures and their solitonic
excitations.  Consider again the DP core structure of Fig.~2(b),
showing the central row of core atoms as well as all of their
first neighbors.  Fig.~2(c) simplifies the picture above it,
replacing the central core atoms with dotted lines indicating
second-neighbor connections of the remaining off-core neighbors.
These are then replaced by a series of lower-case letters that
indicate the sequence of directions of these dashed lines (as
viewed in two dimensions, from the viewpoint of
the black and white atoms):  `u' and `n' indicate
`up' and `down', while `d', `q', `p', and `b' indicate
`upper-right', `lower-right', `lower-left', and `upper-left',
respectively (mnemonically referring to the position of the
typographic stem of the letter).  Thus, the structure of Fig.~1(b)
or 2(b) becomes ``...upnbupnb...'' while that of Fig.~1(a) or 2(a)
would be ``...nbnb...''

As mentioned earlier, the DP structure breaks two symmetries (mirror and
single-period translational symmetries), and has four equivalent ground
states (``dnqu'', ``qudn'', ``pnbu'', ``bupn'') related to each other by
(110) mirrors and by single-cell translations.  We first consider the
antiphase defect that occurs at a translational domain boundary between
core segments; we shall refer to this as a ``phase-switching defect''
(PSD).  The PSD corresponds to a sequence of the form
``...bupnbnbu...'' [Fig.~3(a)] or ``...bupupnbu...''
(or their mirror images).  As can be seen in
Fig.~3(a), a PSD can be regarded as a short segment of the SP structure
inserted into the DP one.  It is free of dangling bonds, and thus is
expected to be a low-energy structural excitation.  Due to the
presence of the stacking fault, the sequences ``...bupnbnbu...''
and ``...bupupnbu...'' are not related by any exact symmetry,
and so will have slightly different energies.

A second class of defects results from a reversal of the mirror
symmetry-breaking.  We shall refer to these as ``direction-switching
defects'' (DSDs); they can be classified by the direction of switching,
among other factors.  Two examples, ``...bup(nu)dnq...'' and
``...qudnnbup...'', are shown in Fig.~3(b) and (c), respectively.  [The
notation `(nu)' indicates a pair of core atoms bonded to the same
out-of-plane atom.]  It turns out to be impossible to build a DSD
without introducing a dangling bond or an overcoordinated atom, so the
DSDs are expected to be more costly than the PSDs.  (The malcoordinated
atoms do not appear in Fig.~3 as they are located just above or below
the plane of the figure.)  Combinations of a DSD and a PSD may also
occur; these also contain a coordination defect.

We have calculated the energies of several of these defects using
the linear-scaling total-energy tight-binding approach.  Supercells
containing up to 768 atoms were employed.  The results are shown
in Table II.  It can be seen that the DSDs do have a higher
energy than the PSDs, as anticipated.  Clearly much work remains
to be done.  One interesting question is that of the
interactions between PSDs and DSDs, and whether the formation of
a PSD-DSD complex would be exothermic.  We also have not yet
studied the mobility barriers for these defects.

\begin{table}
\caption{Energies of various defects, given in eV.  The notation,
described in the text, specifies the entire cell.  Energies for the DSDs
are given for a matched pair of defects.}
\begin{tabular}{llc}
PSD	&bupnbnbupn &0.42	\\
PSD	&bupupnbupn &0.35	\\
DSD+DSD	&bup(nu)dnqudnnbupn	&1.30	\\
DSD+DSD	&bu(pq)udnqudpnbupn	&1.37	\\
\end{tabular}
\end{table}

Finally, we turn to a discussion of kink structures, whose mobility
ultimately determines the mobility of the dislocation as a whole.
Because there are four degenerate core structures to choose between on
each side of the kink, there should be at least 16 distinct kinks.
However, each of these is paired with another into which it can be
converted by displacing the center of the kink by one lattice constant
along the dislocation.  (Using a ``/'' to denote the kink, one such pair
would be ``...qudnq/bupn...'' $\longleftrightarrow$
``...quq/bnbupn...''.)  Thus, we may distinguish 8 topologically
distinct families of kinks.  Furthermore, most of these families may be
classified as ``kink-defect complexes'' incorporating either a DSD, or
PSD, or both, which may or may not be energetically bound to the kink.
Those including a DSD will retain a malcoordinated atom, and will have
no reversal of the mirror symmetry-breaking across the kink; those not
including a DSD will be fully reconstructed and will show a reversal of
the mirror symmetry-breaking.  An example of the latter kind is the kink
``...udnq/bupn..''  shown in Fig.~3(d).  Presumably the free energies of
formation and migration of such kinks are the key quantities determining
the mobility of the 90$^\circ$ dislocation in Si.

In summary, we have proposed a new period-doubled structure for the
the 90$^\circ$ partial dislocation in silicon.  The new DP structure
is predicted to be lower in energy than the SP structure that has
been commonly accepted until now.  Thus, we suggest that it may be
appropriate to reconsider the interpretation of previous experimental
work in view of the proposed DP structure.  As regards the theoretical
work, it is clearly now a high priority to investigate in detail the
structure and energetics of defect and kink structures associated with
the new core reconstruction.

This work was supported by NSF Grants DMR-91-15342 and DMR-96-13648.
One of us (J.B.) wishes to thank N.~Marzari for assistance with the
LDA calculations.


\end{document}